\documentclass[aps,twocolumn,showpacs,superscriptaddress,amsmath]{revtex4}
\include{ams}
\usepackage{amsmath,amssymb,amsthm}
\usepackage{dcolumn}
\usepackage{bm}

\usepackage{graphicx}
\usepackage{epsfig}
\newcommand{\be}{\begin{equation}}
\newcommand{\ee}{\end{equation}}
\newcommand{\bea}{\begin{eqnarray}}
\newcommand{\eea}{\end{eqnarray}}

\begin{document}
\title{Quantum non-equilibrium approach for fast electron transport in
open systems: photosynthetic reaction centers}
\author{M. Pudlak}
\affiliation{Institute of Experimental
Physics, Slovak Academy of Sciences,04353 Kosice,
Slovak Republic}
\author{K.N. Pichugin}
\affiliation{Kirensky Institute of Physics, Akademgorodok 50/38, 660036,
Krasnoyarsk, Russia}
\author{R.G. Nazmitdinov}
\affiliation{Departament de F{\'\i}sica,
Universitat de les Illes Balears, E-07122 Palma de Mallorca, Spain}
\affiliation{Bogoliubov Laboratory of Theoretical Physics,
Joint Institute for Nuclear Research, 141980 Dubna, Russia}
\author{R. Pincak}
\affiliation{Institute of Experimental
Physics, Slovak Academy of Sciences,04353 Kosice,
Slovak Republic}

\begin{abstract}
Creation of electron or exciton by external fields in a system  with
initially statistically independent unrelaxed vibrational modes
leads to an initial condition term. The contribution of this term
in the time convolution generalized master equation approach
is studied in second order of the perturbation theory in the path integral formalism.
The developed approach, applied for analysis of dynamics in the
photosynthetic reaction center, exhibits the key role of the
initial condition terms at the primary stage of electron transfer.
\end{abstract}
\date{\today}
\maketitle

\section{Introduction}
The effect of environment on transport properties of quantum systems
is a highly topical problem in atomic, nuclear and condensed matter
physics. Markovian approaches have been successfully used to study
various phenomena in open systems, when the past memory of the system
is neglected \cite{Weiss}.
The advent in ultrafast laser-pulse technology \cite{ult},
quantum information processing \cite{bar}, synthesis
of new superheavy elements in cold and hot-fusion
reactions \cite{og} are requiring a resolution of quantum dynamics,
when a system is far from equilibrium.

Analogous processes can be found in biological systems \cite{Vault}
as well as in nanoscale devices \cite{Datta}. Although there are
wide structural and functional differences, the laws, that govern
quantum solar energy conversion to chemical energy or electricity in
biological systems and semiconductor solar cells, share many
similarities. In these systems the conversion processes proceed from
the creation of electron-hole pairs (exitons) by a photon of light,
followed by charge separation to produce the required high-energy
product. The efficiency of solar cells may be increased due enhanced
multiple exciton production in semiconductor quantum dots, which is
essentially created extremely fast upon absorption of high-energy
photons \cite{noz}.
On the other hand, the photosynthetic reaction center (RC)
of bacteria provides an interesting system
for studying a high-efficiency electron transfer in an organized
molecular complex. The RC is a special pigment-protein complex, that functions
as a photochemical trap. In such systems,
after excitation the electron transfer is so fast that there can
exist unrelaxed vibrational modes in the primary stage of electron
transfer.

 In a  most studied case  of a purple bacteria
the RC is composed of three protein subunits called L, M and H \cite{2,3}.
According to experimental facts, the protein H does participate in the
electron transfer (ET). All molecules  (cofactors) involved in the ET
are non-covalently bound to subunits L and M in two chains.
Both chains of cofactors start at
the bacteriochlorophyll dimer (P) which interacts with both subunits
L and M. The dimer plays the role of the donor of an electron
(a weakly bound exciton) at the photon absorption.
Cofactors in the subunit L are accessory bacteriochlorophyll
($\mathrm{B}_L$), bacteriopheophytin ($\mathrm{H}_L$) and quinone
($\mathrm{Q}_L$). Identically, in the M subunit there are the
accessory bacteriochlorophyll ($\mathrm{B}_M$), bacteriopheophytin
($\mathrm{H}_M$) and quinone ($\mathrm{Q}_M$).
The cofactors serve as donor-acceptor pairs in the ET.
The arrangement of cofactors shows the local twofold symmetry which is almost
perfect with the respect to the dimer. The part of the L subunit involved in
the ET can be superimposed onto the corresponding part of the M subunit by
a rotation of almost exactly $180^{\circ}$  (for more details on
structural arrangement see \cite{4}).

In spite of the structural symmetry of the two chains of cofactors,
it appears that the RC is functionally highly asymmetric.
In the primary charge transfer an
electron is transferred from the photoexcited dimer P, the
starting point for a series of electron transfer reactions across
the membrane, to the cofactors on subunit L, to
$\mathrm{B}_L$, $\mathrm{H}_L$, $\mathrm{Q}_L$, and
$\mathrm{Q}_M$ \cite{mar,hof}.
On the other hand, the chain located
on subunit M is inactive in the ET. The highly asymmetric
functionality, however, can be decreased by amino acid mutations
or cofactor modification \cite{Heller}.
If the direct ET between subunits L and M is not allowed and electron
cannot escape from the system then it was shown that the different
stochastic fluctuations in the energy of subunits
and the interaction between subunits on these two ways may
cause the transient asymmetric electron
distribution at L and M branches during relaxation to the steady state \cite{Pincak}.
However, due to  fast electron transfer the memory effects should be important at
the primary processes in the photosynthesis.
The major goal of the present paper is to elucidate
the effect of the initial condition terms on electron transfer
in a system with initial conditions being far from equilibrium.

The content of the paper is as follows. In Sec. II we
outline the derivation of the generalized master equation
adapted for physics of the RC.
In Sec. III we use this equation to analyse the contribution of
the initial condition terms on electron transfer in three-site model
of the RC. A brief summary is presented in Sec.IV. In Appendix A
there are details of the GME equation derivation.

\section{Master Equation for Reaction Center}

Formally, an exact generalized master
equation (GME) which describes the electron transfer processes in systems
with dissipation can be constructed by means of
the projection operator techniques \cite{Nak,Zwanzig}.
To be specific let us consider a system in which an
electron has $N$ accessible sites embedded in a medium.
Such a system is described by the Hamiltonian
\be
H=H_0+V ,
\ee
where
$H_0=\sum_{n=1}^{N}|n\rangle[\varepsilon_n-i\Gamma_n+H_n^{v}]\langle
n|$. Here, $|n\rangle$ is the electron state with energy
$\varepsilon_n$.  The
parameter $\hbar/2\Gamma_n$ characterizes the electron lifetime
at site $|n\rangle$. It may originate, for example,
from a nonradiative internal conversion or a recombination process.
The term $H_n^{v}$ describes a medium (a solvent)
consisting of harmonic oscillators
\begin{eqnarray}
\label{3th4}
H_n^{v}=\sum_a\bigg\{\frac{p_\alpha^{2}}{2m_\alpha}+
\frac{1}{2}m_\alpha\omega_\alpha^{2}(q_\alpha-d_{n\alpha})^{2}\bigg\}\;.
\end{eqnarray}
Here, $d_{n\alpha}$ is the
equilibrium configuration of the $\alpha$th oscillator, which depends
on the electronic state $|n\rangle$. The interaction
$V=\sum_{n,m=1}^{N}V_{nm}|n\rangle \langle m| (n\neq m)$ couples different
sites.

The GME formulation of an electron (an exciton) interacting with
vibrational modes (phonons) starts from the Liouville
equation for a density operator $\rho (t)$ \cite{Weiss}
\begin{equation}
i\frac{\partial}{\partial t}\rho (t)=\frac{1}{\hbar}[H\rho (t)-\rho
(t) H^{+}]\equiv L \rho (t)\;.
\end{equation}
At this stage the projector operator technique allows to
avoid a knowledge of a complete information upon  a system under
consideration. The projector operator contracts the full information
about the system to the relevant one. Our prime interest is the
information about the electron localization and the irrelevant
information is a particular vibrational state excited in the system.
In virtue of this technique \cite{Nak,Zwanzig} one obtains
\bea
\label{tder}
&&\frac{\partial}{\partial t}D\rho (t)= -iDLD\rho (t)-\\
&&-\int_{0}^{t}DL\exp[-i(1-D)L\tau](1-D)LD\rho (t-\tau)d\tau- \nonumber\\
&&-iDL\exp[-i(1-D)Lt](1-D)\rho (0)\nonumber
\eea
for the relevant part $D\rho(t)$ of the total density $\rho(t)$. Here
$D=D^{2}$ is an arbitrary linear projection operator which can be
used in the form \cite{Pudlak}
\be
\label{da}
DA=\sum_{n}Tr\left(\  |n\rangle \langle n|\ A \ \right) \rho_{n}|n
\rangle \langle n|\;.
\ee
The total trace $Tr=Tr_{e}Tr_{Q}$ is a product of traces
of the electronic ($Tr_{e}$)
and the vibrational ($Tr_{Q}$) subsystems;
$\rho_{n}=\exp(-\beta H_{n}^{v})/Tr_{Q}[\exp(-\beta H_{n}^{v})]$ is a
density operator  for a vibrational subsystem,
when an electron is localized at a site $|n\rangle$
($\beta=1/k_{B}T$). Often, the projector operator is chosen in such
a way that the initial state $(1-D)\rho(0)$ is disregarded \cite{Weiss}.
This approximation is valid when electron and phonon subsystems are
initially in equilibrium. Evidently, that at the primary stage
of the electron transfer this term may  influence the
electron pathway in the RC. The questions arise about the time-scale
of such influence in the system under consideration and
how this influence would affect the ET.

To calculate the initial condition terms we have to specify $\rho(0)$.
The photon absorption by the dimer results in the transition of an electron
 from the ground to the
excited state of the dimer (say, the excited state $|1\rangle$).
Before the excitation the electronic subsystem is in the thermal
equilibrium with the vibrational subsystem which consists in
vibrational modes of the dimer and the protein subunits. Due to the
fast electron transfer to molecules located in L(M)-branches of the
RC the time is too short to establish a thermal equilibrium between
the  vibrational subsystem and a new electronic state. On the other
hand, the vibrational subsystem (the bath) is in the thermal
equilibrium with the electronic ground state of dimer.
Thus, we have
\be
\label{im}
\rho(0)=\rho_{0}\otimes
\left(\sum_{kl} \rho_{kl}^{e}(0)|k\rangle\langle l|\right)
\ee
where $\rho_{0}=\exp(-\beta H_{0}^{v})/Tr_{Q}[\exp(-\beta
H_{0}^{v})]$ is a density operator for a vibrational subsystem, when
an electron is at the ground state. Here
$\rho_{kl}^{e}(0)$=$Tr_{Q}(\langle k|\rho (0)|l\rangle )$ is an
electronic part of the density matrix. We suppose that an electron is
initially localized on the first molecule:
i) $\rho_{11}^{e}(0)=1$, $\rho_{nn}^{e}(0)=0,n=2,N$;
ii)non-diagonal initial density
matrix elements are $\rho_{kl}^{e}(0)=0$.
We also assume that after the excitation
the electron transfer is so fast that the initial vibrational
density is not affected.
Similar assumptions for the construction of the initial state
$\rho(0)$ have been used for analysis of the energy transfer
dynamics in a model of a donor-acceptor pair \cite{ McCutcheon}.
In principle, the initial conditions may be calculated  within
a scheme proposed to include system-bath correlations after the interaction
with optical pulses \cite{ap}.  However, this problem requires a dedicated
study itself in order to distinguish different time scales and is beyond the scope
of the present consideration. The main objective here is to gain insights into
the role of the initial condition terms on the ET for a few typical cases.

With the aid of Eq.(\ref{da}) one obtains
\be
\label{imd}
D \rho(0)=\rho_{1}\otimes|1\rangle\langle 1|\;,
\ee
which leads to the initial state
\be
(1-D)\rho(0)=(\rho_{0}-\rho_{1})\otimes|1\rangle\langle 1|.
\ee
 Evidently, if the condition $\rho_{0}\simeq\rho_{1}$
is not fulfilled, one must take into account
the initial state in Eq.(\ref{tder}).

Substituting Eq.(\ref{da}) in Eq.(\ref{tder}), one
obtains the GME
\bea
\label{gme1}
\partial_tP_n(t)& = &-\frac{2\Gamma_n}{\hbar}P_n(t)-
\sum_{m(\neq n)}\int_{0}^{t}[Re W_{nm}(t-\tau)P_n(\tau)-\nonumber\\
&&-Re W_{mn}(t-\tau)P_m(\tau)]d\tau + I_{n}(t)
\eea
for site occupation probabilities
\be
P_{n}(t)=Tr\left(\  |n\rangle \langle n|\ \rho (t)\
\right)=\rho_{nn}(t)\;.
\ee
The first term in the r.h.s. of  Eq.(\ref{gme1})
is associated with the  probability for an electron to escape
from the system via an additional channel.

In the Born approximation, the memory function $W_{mn}(t)$ can
be expressed in the form
\bea
\label{wmn}
&&W_{mn}(t)= 2
\frac{|V_{mn}|^{2}}{\hbar^{2}}\exp\left(-\frac{\Gamma_{m}+
\Gamma_{n}}{\hbar}t\right)\times\\
&&\exp\left(i\frac{\varepsilon_{m}-\varepsilon_{n}}{\hbar}t\right)
\exp\bigg(\sum_{\alpha}\frac{E_{mn}^{\alpha}}{\hbar \omega_{\alpha}}\times\nonumber\\
&&[(\bar{n}_{\alpha}+1)e^{-i\omega_{\alpha}t}+
\bar{n}_{\alpha}e^{i\omega_{\alpha}t}-(2\bar{n}_{\alpha}+1)]\bigg).\nonumber
\eea
Here
$\bar{n}_{\alpha}=[\exp(\hbar\omega_{\alpha}/k_{B}T)-1]^{-1}$
is a thermal population of the $\alpha$th vibrational mode and
\begin{equation}
\label{ren}
E_{mn}^{\alpha}=\frac{1}{2}m_{\alpha}\omega_{\alpha}^{2}(d_{m\alpha}-d_{n\alpha})^{2}
\end{equation}
is the reorganization energy of the $\alpha$th vibrational mode, when
an electron moves from state $|m\rangle$ to state $|n\rangle$.

In second order of the perturbation theory, one obtains from
Eq.(\ref{tder}) the initial condition term (IT) in the form
(see also \cite{Capek,Golosov1})

\be
I_{n}(t)=I_{n}^{(1)}(t)+I_{n}^{(2)}(t)
\ee
where the first order term is
\be
\label{foit}
 I_{n}^{(1)}(t) = -iTr\left(\  |n\rangle \langle n|
\mathcal{L}e^{-iL_{0}t}(1-D)\rho (0)\right)
\ee
and the second order term has the form
\be
\label{stit}
I_{n}^{(2)}(t) =
-\int_{0}^{t}d\tau \Phi(t,\tau).
\ee
Here,
\be
\label{prit}
\Phi(t,\tau)=\ Tr\left(|n\rangle \langle n|
\mathcal{L}e^{-iL_{0}(t-\tau)}
\mathcal{L}e^{-iL_{0}\tau}(1-D)\rho (0)\right)
\ee
and
\be
\label{eq2}
\mathcal{L}A=\frac{1}{\hbar}[V,A], \quad
L_{0}A=\frac{1}{\hbar}[H_{0},A]\;.
\ee

The choice of the initial conditions
Eqs.(\ref{im}),(\ref{imd}), and
Eqs.(\ref{eq2}) yield
\be
\label{foz}
I_{n}^{(1)}(t)=0\;.
\ee
The ITs fulfil the general identity which transforms in the considered case
to
\be
\label{rav}
\sum_{n} I_{n}(t)=0 \Rightarrow I_{1}^{(2)}(t)=-\sum_{n=2}^{N} I_{n}^{(2)}(t).
\ee
Evidently, the definition of $I_{n}^{(2)}(t)$ is crucial for
the calculation of the GME (\ref{gme1}).
 For $I_{n}^{(2)}(t)$ we obtain the
following expression (see details in Appendix)
\bea
\label{main}
&&I_n^{(2)}(t)=Re \int_0^td\tau W_{1n}(t-\tau){\cal B}, \ \ n=2,....N  \\
&&{\cal B}=\exp\bigg[2{\rm i}
\sum_{\alpha=h,v}S_{10}^{\alpha} {\cal M}_{\alpha} \bigg] -1\nonumber
\eea
where the variable $S_{10}^{\alpha}$
(see Eq.(\ref{smn}))
is determined by the reorganization energy of the $\alpha$ vibrational
mode $E_{10}^\alpha$, Eq.(\ref{ren}), when electron moves from the
ground state $|0\rangle$ to the excited state $|1\rangle $ of the dimer P.
As a check, using the same techniques, we have calculated $ I_{1}^{(2)}(t)$
and obtained the fulfillment of Eq.(\ref{rav}).

Note that if one considers the contribution of the nondiagonal
density matrix elements, the first order term contributes
to the GME as well \cite{Jang,Jang1}.

\section{Model of primary stage of electron transfer in a reaction center}

To demonstrate the viability and utility of our approach,
we consider the electron transfer in RCs within a three-site model.
In the RC, after photon absorption at the bacteriochlorophyll dimer
molecules (molecule 1) the electron transfer may
occur either through the M-side bacteriochlorophyll (molecule 2)
or the L-side bacteriochlorophyll (molecule 3)  accessors.  We
assume that there are two non-zero coupling
terms $V_{12}$ and $V_{13}$ and a forbidden direct electron
transfer between molecules 2 and 3 ($V_{23}=0$). The two possible
electron transfer pathways are related by the $C_{2}$ symmetry axis.
The pathway symmetry is broken by differences in amino
acid around the donor-acceptor pairs in the different branches.
These differences inhibit a charge separation through the M-branch in
the wild type RC. The preference of the pathway $(1\rightarrow3)$
relative to the pathway  $(1\rightarrow2)$ is assumed to be governed by
the energy differences and by the different couplings between donor-acceptor
pairs in the branches. Indeed, experimental and theoretical estimates
for the energy differences
provide the following figures \cite{Holten,Alden}:
$\varepsilon_{1}-\varepsilon_{3}\approx
0.05-0.1$,\;\;$\varepsilon_{2}-\varepsilon_{1}\approx
0.1-0.2$ (eV).

The Hamiltonian $H_{0}^{v}$ determines the vibrational state,
the equilibrium position of the oscillatory
mode, and, consequently, the density operator for the vibrational subsystem,
when an electron is in the base state of the
donor molecule before the excitation.
The phonon bath is described
by two vibrational modes, high
frequency mode $\omega_{h}$ and low frequency $\omega_{v}$. The
low-frequency mode characterizes the exterior medium phonon mode and
the high-frequency mode describes the molecular vibrational modes of
the donor and the acceptor centers \cite{Jortner,Tanaka}.
We take into account a relaxation time for
the vibrational mode
$\omega_{\alpha}\rightarrow \omega_{\alpha}-i/\tau_{\alpha}$
with the aid of a phenomenological parameter $\tau_{\alpha}$
(see also \cite{Bialek}).
As a result, for numerical analysis the variables (\ref{nal}), (\ref{mal})
 are considered in the form
\bea
 {\cal N}_{\alpha}=(2\bar{n}_{\alpha}+1)
[1-e^{-t/\tau_{p\alpha }}\cos(\omega_{\alpha}t)]
+{\rm i}e^{-t/\tau_{p\alpha}}\sin(\omega_{\alpha}t)\nonumber \eea \bea
{\cal M}_{\alpha}=\exp(-t/\tau_{p\alpha})\sin(\omega_{\alpha}t)
-\exp(-\tau/\tau_{p\alpha})\sin(\omega_{\alpha}\tau)\nonumber
\eea
To illustrate the contribution of the ITs in the ET, we
use parameters that may elucidate in the Markovian
approximation the observed L-side experimental kinetics of wild-type (WT) RCs of
{\it Rb. sphaeroides} ~\cite{Tanaka,Bixon}. Note that the probability of
the M-side electron transfer was excluded in such considerations.

In partiuclar, the following set of parameters (set I) is used to characterize the
electron transfer in a wild type RC via the L branch
(the pathway $1\rightarrow 3$):
$V_{13}=7$ meV, $\hbar\omega_{h}=187$ meV, $\hbar\omega_{v}=12.5$ meV,
$S_{13}^{h}=0.5$, $S_{13}^{v}=8$ \cite{Tanaka}.
The same set can be used for the electron transfer via the
M branch (the pathway $1\rightarrow 2$), taking into account
that there is a two-fold symmetry.
To provide the asymmetry in the electron transfer via the
two pathways we consider:
$\varepsilon_{3}-\varepsilon_{1}=-56$ meV
and  $\varepsilon_{2}-\varepsilon_{1}=110$
meV (see discussion in \cite{Pincak}).
This choice forms the set II.

\begin{figure}
\label{fig1}
\begin{center}
\includegraphics[width=8cm]{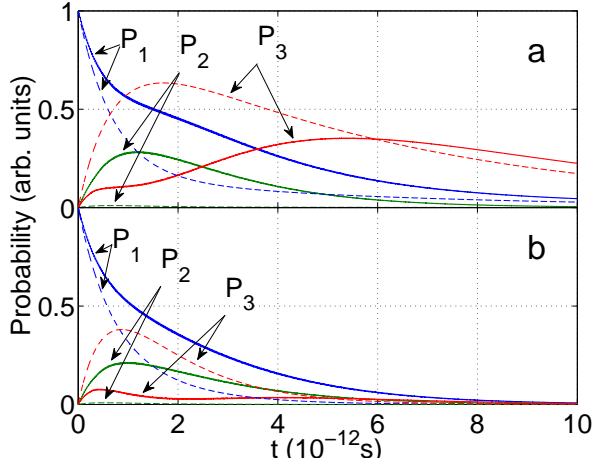}
\caption{(Color online) Time dependence of the site occupation probabilities
$P_{i}(t)$. Solid (dashed) lines are used for results with (without)
the ITs. The basic parameters are used for the both panels.
The specific parameters are:
$S_{10}^{h}=1$,$S_{10}^{v}=0$;
a)$2\Gamma_{2}/\hbar = 2\Gamma_{3}/\hbar
=(5ps)^{-1}$;
b)$2\Gamma_{2}/\hbar = 2\Gamma_{3}/\hbar
=(1ps)^{-1}$}
\end{center}
\end{figure}

Evidently, that the RC is an open system which interacts
with another part of overall system. This part can be assumed to
have a quasi-continuum spectrum. To mimic the realistic situation we introduce
sink parameters $\Gamma_{i}/\hbar$ $(i=1,3)$ which characterize the electron transfer
to another part of the overall system with a quasi-continuum spectrum.
This is an effective approach to describe resonance scattering 
phenomena in open systems with a weak coupling to the environment 
(see, for example, \cite{bil}).
When an electron is transferred to this subsystem, the backward
electron transfer can be neglected. 
The primary charge separation step $(1\rightarrow3)$
occurs in purple bacterial RCs  with a lifetime $\sim 2$ps at room
temperature. The next step of electron transfer occurs
in the time scale $\sim 1$ps \cite{Artl}.
In the numerical calculations we
have used the following values of sink parameters:
$2\Gamma_{1}/\hbar=(250ps)^{-1}$\cite{Laporte},
$2\Gamma_{2}/\hbar =2\Gamma_{3}/\hbar=(5ps)^{-1}$ and
$(1ps)^{-1}$. The parameter $2\Gamma_{1}/\hbar$
characterizes the decay of the system to the ground state, while
$2\Gamma_{2,3}/\hbar$ are associated with the electron transfer to the
next molecules (subsystems) which are beyond the scope
of the present analysis.
The flow direction also depends on the parameter
$S_{10}^{h}$ which characterizes the amount of energy
stored in unrelaxed high frequency vibrational modes.
 The scale reorganization
constant values are chosen as $S_{10}^{h}=0.5 $ and $S_{10}^{h}=1$.
We take $\tau_{\alpha}\simeq 3.5ps$ for the
lattice relaxation time, in accordance with the observation
that a vibrational mode relaxation time is of order
a few ps \cite{Sumi}. The set III consists of $2\Gamma_{1}/\hbar$ and $\tau$.
The sets I, II, III form the basic parameters of calculations
of occupation probabilities at 300K, shown at Figs.1,2.

Let us consider a case when the lattice relaxation
time is smaller than ones defined by the  sink parameters
(see Fig.1a). The system has time to
achieve the regime when the impact of the unrelaxed phonon mode on the
occupation probabilities is easing. This imposes the impact of
the ITs upon the  electron flow direction. The electron flow direction
is similar to the one defined by the theory without
the ITs.  At the early stage of electron transfer the ITs  have, however,
a strong influence on the quantum yields  of the electron flow
through different branches. The theory without the ITs predicts
that the M branch is inactive. The ITs activates
the  electron escape through the branch M (molecule 2).

The energy stored in the unrelaxed
phonon modes is large ($S_{10}^{h}=1$) in this case. The parameters
$\hbar/2\Gamma_{2}=\hbar/2\Gamma_{3}=5ps$ are large enough in
comparison with the phonon relaxation time $\tau_{p}=3.5ps$. The
system evolves in time to the regime where the ITs influence
decreases on the electron transfer (the system begins to forget the
initial conditions). Therefore, the flow direction is the same as in
the case without the ITs.
If the sink parameters are smaller than the phonon relaxation time
(see Fig.1b) the ITs change the favourite partway for the electron transfer:
the dominance of the pathway $(1\rightarrow 3)$ (without the ITs) is replaced
by the dominance of the pathway $(1\rightarrow 2)$ (with the ITs).

Let us consider the regime, when the unrelaxed phonon modes of the
medium are taken into account($S_{10}^{v}=2$) (Fig.2). Such amount
of the unrelaxed medium modes has no a strong impact on the main
characteristic of electron transfer. The medium modes can not store
large amount of the unrelaxed energy. Indeed,  the transition of the
system to the excited state affects rather the rearrangement of atom
positions in the donor-molecule and, thus, has no a strong impact on
the medium atom positions. The ITs increase the probability for
electron transfer via the molecule 2 in comparison to the case
without the ITs. The electron lifetime in the system is still short
enough not to forget the initial condition. With the
decrease of unrelaxed phonon energy in the system (compare
Figs.2a,b) the importance of the ITs is also decreasing.

\begin{figure}
\begin{center}
\includegraphics[width=8cm]{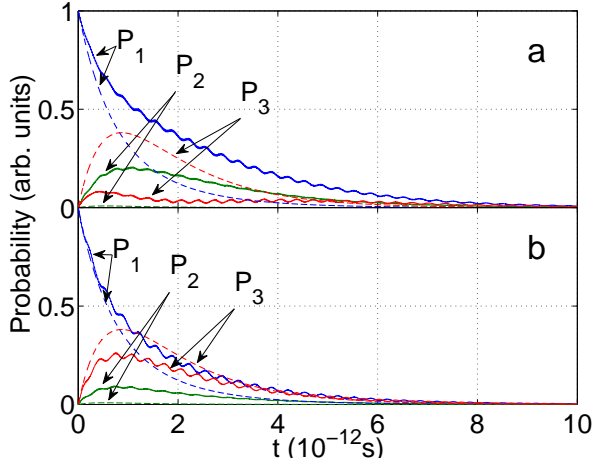}
\caption{(Color online) The basic parameters are the same as in
Fig.1b. The specific parameters for this figure are: $S_{10}^{v}=2$.
a)$S_{10}^{h}=1$; b)$S_{10}^{h}=0.5$.}
\end{center}
\end{figure}

Another important ingredient is the dependence of the results on the
ratio of the phonon relaxation time and the time associated with the
interaction V. All results discussed above are related to the
situation when $\tau > \hbar/V_{1m}$. In this case
the electron transitions are fast, while the phonon relaxation 
is a slow process. The  ITs will contribute to the ET depending
on the amount of energy stored in the unrelaxed
vibrational modes, in accordance with 
the discussion above. 
On the other hand, if $\tau \ll \hbar/V_{1m}$
the ITs produce a marginal effect, and the initial conditions are
forgotten relatively quickly.

\section{Summary}
In conclusion, we suggest the microscopic approach
to study the effect of the ITs on the electron transfer
in a system with initial conditions being far from equilibrium.
The IT impact depends mainly on the amount of energy
stored in the initially unrelaxed phonon modes
 and also on the lifetime of the electron in the system.
If the electron lifetime is much longer than the phonon relaxation time,
the ITs do not affect the quantum yields of electron transfer
via possible pathways.
In systems, where this condition is not fulfilled,
the ITs can cause the electron transfer via
channels which are closed in the case without the ITs.

\appendix

\section{Initial condition terms}
\label{app1}

Let us employ the initial conditions (\ref{im}),(\ref{imd}), in the definition of
the second order of the initial term (\ref{stit}),(\ref{prit}),
\bea
I_{n}^{(2)}(t)&=&
\int_{0}^{t}d\tau \ Tr\bigg{(}|n\rangle \langle n|
\mathcal{L}e^{-iL_{0}(t-\tau)}\times\\
&\times&\mathcal{L}e^{-iL_{0}\tau}(\rho_{1}-\rho_{0})|1\rangle
\langle 1|\ \bigg{)}\;,\;n=2,...N.
\nonumber
\eea
With the aid of Eqs.(\ref{eq2}) we obtain
\bea \label{in2}
I_{n}^{(2)}(t) &=&-2Re \frac{|V_{1n}|^{2}}{\hbar^{2}}\int_{0}^{t}d\tau\times\\
&\times&\exp\left(-\frac{i}{\hbar}(\varepsilon_{n}-\varepsilon_{1})(t-\tau)\right)\times\nonumber\\
&\times& \exp\left(-\frac{1}{\hbar}(\Gamma_{n}+\Gamma_{1})(t-\tau)\right) \times\nonumber \\
&\times&Tr_{Q} \left(e^{\frac{i}{\hbar}H_{1}^{v}t}e^{-\frac{i}{\hbar}H_{n}^{v}
(t-\tau)}e^{-\frac{i}{\hbar}H_{1}^{v}\tau}(\rho_{1}-\rho_{0})\right)\nonumber
\eea

To proceed further it is necessary to calculate the trace over the
environmental (vibrational) degrees of freedom, which has, in general,
the following form

\bea
{\cal I}=Tr_{Q} \left(e^{\frac{i}{\hbar}H_{m}^{v}t}e^{-\frac{i}{\hbar}H_{n}^{v}
(t-\tau)}e^{-\frac{i}{\hbar}H_{m}^{v}\tau}\rho_{k}\right)
\eea

In the path integral formalism this trace can be written as

\bea
\label{core}
{\cal I}&=&\sum_{\nu_{\alpha}}\int_{-\infty}^{\infty}dq
\int_{-\infty}^{\infty}dx_{1}\times\nonumber\\
&\times& \int_{-\infty}^{\infty}dy
\int_{-\infty}^{\infty}dy_{1}
\langle k\nu_{\alpha}|q\rangle\\
&\times&\langle
q|\exp\left(i\frac{H_{m}^{v}}{\hbar}t\right)|x_{1}\rangle
\langle x_{1}|\exp\left(-i\frac{H_{n}^{v}}{\hbar}(t-\tau)\right)|y\rangle\nonumber\\
&\times&\langle y|
\exp\left(-i\frac{H_{m}^{v}}{\hbar}\tau\right)|y_{1}\rangle \langle
y_{1}|\exp\left(-\beta H_{k}^{v}\right)|k\nu_{\alpha}\rangle /{\cal Z}.\nonumber
\eea
Here, ${\cal Z}=Tr_{Q}\left[\exp(-\beta\ H_{k}^{v}\right)]$ and
a vibrational state is
 \bea
&&\langle q|k \nu_{\alpha}\rangle =\chi_{\nu_{\alpha}}(z)= {\cal
N}\times
H_{\nu_{\alpha}}(z)\exp\left(-z^2/2\right)\nonumber\\
&& {\cal N}= \left(\frac{1} {\ell_0
\pi^{1/2}2^{\nu_{\alpha}}\nu_{\alpha}!}\right)^{1/2}, \quad z=
(q-d_{k\alpha})/ \ell_0\;, \nonumber
\eea
where $\ell_0=\sqrt{\hbar/m\omega_{\alpha}}$ is the oscillator length,
and $H_{\nu_{\alpha}}$ is Hermitian polynomial of
order $\nu_{\alpha}$.

In order to obtain an analytical result
we suggest to calculate this expression in the integral
path formalism \cite{Feynman}. As a result, Eq.(\ref{core}) takes
the following form
\bea
&&{\cal I}=\sum_{\nu_{\alpha}}\int_{-\infty}^{\infty}dq
\int_{-\infty}^{\infty}dx_{1} \int_{-\infty}^{\infty}dy
\int_{-\infty}^{\infty}dy_{1}
K_{m}(q,-t;x_{1})\times\nonumber\\
&& K_{n}(x_{1},t-\tau;y) K_{m}(y,\tau;y_{1})\chi_{\nu_{\alpha}}(y_{1}-d_{k\alpha})\times\\
&&\exp[-\beta(\nu_{\alpha}+1/2)\hbar\omega_{\alpha}]
\chi_{\nu_{\alpha}}(q-d_{k\alpha}),\nonumber
\eea
where
\bea
\label{core1}
&&K_{i}(x,t;y)\equiv \langle x|\exp(-iH_{i}^{v}t)|y\rangle =\nonumber\\
&&=\sqrt{\frac{1}{2i\pi \ell_0^2 \sin\omega_{\alpha} t}}
\exp\bigg{\{}i\frac{{\cal F}_0}{2\ell_0^2\sin\omega_{\alpha} t}\bigg{\}}\\
\label{fn}
&&{\cal F}_0= \cos\omega_{\alpha} t[(x-d_{i\alpha})^{2}
+(y-d_{i\alpha})^{2}]\nonumber\\
&&-2(x-d_{i\alpha})(y-d_{i\alpha})\;.
\eea

In virtue of the equation
\bea
&&\sum_{m}\chi_{m}(x-d_{0})\exp[-\beta(m+1/2)\hbar\omega_{\alpha}]\chi_{m}(y-d_{0})=\nonumber\\
&&=\sqrt{\frac{1}{2\pi \ell_0^2 \sinh(\beta\hbar\omega_{\alpha})}}
\exp\bigg{\{}-\frac{{\cal F}_1}{2\ell_0^2\sinh(\beta\hbar\omega_{\alpha})}\bigg{\}}\\
&&{\cal F}_1=\cosh(\beta\hbar\omega_{\alpha})[(x-d_{0})^{2}+(y-d_{0})^{2}]\nonumber\\
&&-2(x-d_{0})(y-d_{0})\nonumber
\eea
we obtain
\bea
\label{mt}
{\cal I}&=&\exp\bigg{\{}-S_{mn}^{\alpha}{\cal N}_{\alpha}\bigg{\}}
\exp\bigg{\{}2iS_{mk}^{\alpha}{\cal M}_{\alpha} \bigg{\}},\\
\label{smn}
S_{mn}^{\alpha}&=&E_{mn}^{\alpha}/\hbar\omega_{\alpha}\;,\\
\label{nal}
{\cal N}_{\alpha}&=&(2\bar{n}_{\alpha}+1)
(1-\cos\omega_{\alpha}(t-\tau))+\\
&+&i\sin\omega_{\alpha}(t-\tau)\;,\nonumber\\
\label{mal} {\cal M}_{\alpha}&=&\sin\omega_{\alpha} t
-\sin\omega_{\alpha}\tau\;.
\eea
Taking into account the
definitions (\ref{in2}), (\ref{wmn}), with the aid of Eq.(\ref{mt}),
for $m=1$ we obtain the result (\ref{main}).

\section*{Acknowledgement}
This work is partly supported by the Slovak Academy of Sciences in
the framework of CEX NANOFLUID, VEGA Grant No. 2/0069/10,
Grant No. FIS2008-00781/FIS (Spain) and RFBR Grants No. 11-02-00086
(Russia).

\end{document}